\documentclass{jpsj2}
\title{
Component Ratios of Independent and Herding Betters in a 
Racetrack Betting Market
}

\author{Shintaro \textsc{Mori}$\thanks{E-mail :
mori@sci.kitasato-u.ac.jp}$ and Masato
\textsc{Hisakado}$^{1}$}

\inst{
Department of Physics, School of Science, Kitasato University
\\ Kitasato 1-15-1, Minami-ku, Sagamihara, Kanagawa 252-0373, Japan \\
$^{1}$
Standard and Poor's, Marunouchi 1-6-5, Chiyoda-ku, Tokyo 100-0005, Japan
}

\abst{
 We study the time series data of the racetrack betting market 
in the Japan Racing Association (JRA). 
As the number of votes $t$ increases, 
the win bet fraction $x(t)$ converges to the final win bet fraction
$x_{f}$. 
We observe the power law $(x(t)-x_{f})^{2} 
\propto t^{-\beta}$ 
with $\beta\simeq 0.488$.
We measure the degree of the completeness of the ordering 
of the horses using an index AR, the horses are ranked according to the size of the win bet fraction. 
AR$(t)$ also obeys the power law and behaves as 
$\mbox{AR}_{f}-\mbox{AR}(t)\propto t^{-\gamma}$ with 
$\gamma\simeq 0.589$, where $\mbox{AR}_{f}$ is the final value of AR.   
We introduce a simple voting model with two types of 
voters--independent and herding. 
Independent voters provide information about the 
winning probability of the horses and herding voters 
decide their votes based on the popularities of the horses.
This model can explain two power laws of the betting process.
The component ratio of the independent voter to the herding voter is 1:3.
}

\kword{
power law, betting market, stochastic process, accuracy ratio, 
efficiency, herding
}

\begin{document}
\maketitle

\section{Introduction}
 Racetrack betting is a simple exercise of gaining a profit or losing
 one's wager. However, 
 one needs to make a decision in the face of  uncertainty,
 and a closer inspection of this decision-making process 
 reveals great complexity and scope. 
 The field has attracted many academics from various 
 disciplines and has become a subject of wider 
 importance \cite{Ziemba}. 
 Compared to the stock or currency 
 exchange markets, 
 racetrack betting is a short-lived and repeated market.
 It is possible to obtain a clearer understanding of aggregated betting behaviour
 and study the market efficiency. 
 One of the main findings of the previous
 studies is the `favorite-longshot bias' in the racetrack betting market 
 \cite{Griffith,Ziemba2}. The final odds are, on average, accurate measures
 of winning, 
 short-odds horses are systematically undervalued, and long-odds horses are 
 systematically overvalued.
 
 From an econophysical perspective, racetrack
 betting is an interesting subject. 
 Park and Dommany have analysed 
 the distribution of the final odds (dividends) of 
 the races organised by
 the Korean Racing Association \cite{Park}. 
 They observed that the distribution of the final odds exhibited the power law behaviour.
 They explained this behaviour on the basis of  the assumptions
 of a rational better who maximizes the expected payoff
 and a multiplicative rule in the estimate of the winning probability. 
 Ichinomiya also observed the power law of final odds
 in the races organised by the Japan Racing Association 
 (JRA) \cite{Ichinomiya} and proposed
 another betting model where the strength of the horse obeys uniform distribution and the 
 complex competition process in the horse race is described by normal distribution. 
 He also assumed that betters exhibit irrational behaviour and bet their money on the horse that appears to 
 be strongest in the race. 
 The authors analysed the uncertainty in the prediction of 
 the racetrack betting market. 
 It is a short-lived and repeated market, and hence the accuracy of the predictions can be estimated. 
 We found a scale-invariant relation between 
 the rank of a racehorse and the result of its victory 
 or defeat in JRA \cite{Mori}. Horses are ranked according to 
 the win bet fractions. 
 In the long-odds region, 
 between the cumulative distribution function of the winning horses  
 $x_{1}$ and that of the losing horses $x_{0}$, a scale-invariant 
 relation $x_{1}\propto x_{0}^{\alpha}$ with $\alpha=1.81$ holds.
 In a betting model  where
 betters display herd-like behaviour (herding better) with only a small amount of information about the strength 
 of the horses and vote on the horses according to  the 
 probabilities that are proportional to the number of the votes, it is possible to show that the scale 
 invariance emerges in a self-organized fashion. 
 The authors also studied another betting model with two types of voters--independent and herding \cite{Hisakado}. 
 Independent voters provide information about the 
 winning probability of the horses to the market. It was  found 
 that a phase transition occurs in the process of information aggregation and that   
 herding voters are responsible for the slow convergence of the win bet fraction. 

 Summarizing these studies,  we are faced with two questions about the
 racetrack betting market.
 The first question is whether the betters are rational or not. 
 The final odds contain an accurate estimate of the winning probability, which means that
 the betters look rational. However, it is improbable that all betters 
 are clever and are able to precisely  estimate the winning probability. In the exchange or stock 
 market, the role of fundamental and chartist-type participants has been discussed \cite{Alfi}.
 The component ratios of the two types of participants change drastically, and thus, the market exhibits a 
 complex behavior. It was also discussed that herding voters 
 increase the accuracy of prediction in a forecasting game, and such herding behaviour
 may be very efficient in aggregating dispersed  private information \cite{Curty}.  
 In the racetrack betting market,
 whether such types of participants exist is an interesting question.       

 The second question is the distribution of final odds, which reflects the winning 
 probability of a horse. Previous studies have proposed the two possibility of potential mechanisms.
 The drawback of Park and Dommany's model  
 is that the true winning probability 
 distribution does not come from the betting or estimating process. 
 It instead comes from the system in which many 
 horses run at the same time  and try to get to the first position after complex competition  
 process. If the multiplicative estimation rule produces the empirical
 distribution of the winning probability, the model can represent the complex competition process. 
 Betters understand how to estimate the winning probability after studying 
 many horse races; the final odds reflect the 
 true winning probability. Although Ichinomiya proposed an interesting 
 mechanism to observe  power law in the final odds, his assumption of 
 irrational betters cannot be accepted. If betters are not rational, it is 
 difficult to understand the efficiency of the market where the win bet 
 fraction coincides with the true winning probability. 
 In this study, we 
 focus on the rationality of the betters. We study the time series data of the win bet
 in the JRA and  the nature of the betters, i.e., whether they 
 are rational or not, and  what type of better exists in the racetrack betting 
 market. With respect to the distribution of the final odds, we only assume that the winning 
 probability has a broad distribution and can be estimated on the basis of the final odds.

The organization of this paper is as follows.
In  \S \ref{Data}, we provide a detailed study of the time series data of the win bet. 
 We have studied the time series data of the win bet odds
 in 2008 of JRA. 
Horses are ranked according to the win bet fraction and  
the receiver operation characteristic (ROC) curve is discussed.
We measure the fluctuation of the win bet fraction $x$ and 
the degree of the completeness of the ordering 
of the horses by an index AR with the progress of the voting.
In \S \ref{Result}, we show the result of the data analysis.
As the number of vote $t$ increases, $x(t)$ converges
to the final values $x_{f}$ very slowly. 
The power law relation $(x(t)-x_{f})^{2}\propto t^{-\beta}$ with
$\beta \simeq 0.488$ holds.
AR$(t)$ also obeys a power law  
$\mbox{AR}_{f}-\mbox{AR}(t)\propto t^{-\gamma}$ with 
$\gamma \simeq 0.589$, where $\mbox{AR}_{f}$ is the final value of AR. 
In  \S \ref{Model}, we introduce a voting model, 
where there are two types of 
voters--independent and  herding.
Using the exponent in the power law convergence of $x(t)$, 
we estimate 
the component ratio of the independent voter to the herding voter is 1:3.
The power law convergence of AR$(t)$ is also observed in this voting model.
Section \ref{Conclusion} is dedicated to the summary.

\section{Racetrack Betting Process}
\label{Data}
 We study the win bet
data of JRA races in 2008. A win bet is the wager that the better lays on 
 the winner of the race. Out of the 3542 in that year,
 we choose  2471 races whose final public win pool (total number of votes) 
$V^{r}$ is in the range of 
 $10^{5}\le V^{r} \le 3\times 10^{5}, r \in \{1,2,\cdots,R=2471 \}$.
 The average value of $V^{r}$ is about $1.89\cdot 10^{5}$. 
 $N^{r}$ horses run in race $r$; $N^{r}$ is in the range pf $7 \le N^{r} \le 18$.
We ignore 102 canceled horses; the total number of horses  
$N\equiv \sum_{r=1}^{R}N^{r}$ is $35719$. The number of the winning horse is 
2472
 (one tie occurs) and is denoted as $N_{1}=2472$.
 The number of the remaining  horses (losing horses) is denoted as 
$N_{0}=N-N_{1}$. 
$K^{r}$ denotes the number of times a  public announcement was made regarding the  
 temporal odds and number of votes in race $r$. $K^{r}$ is
in the range of $13\le K^{r} \le 217$, and the total number of announcements 
is $K \equiv \sum_{r=1}^{R}K^{r}\simeq 2.0\times 10^{5}$. On an average, 
announcements were made eighty times up till the start of the races.
The time from the announcement to the race entry time (start of the race) 
in minutes is denoted 
as $T^{r}_{k}$.
 We denote the temporal odds of the $i$th horse in race $r$ at the $k$th announcement  
as $O_{i,k}^{r}$; the public win pool as $V^{r}_{k}$.
$V^{r}_{K^{r}}=V^{r}$ holds. 
$I^{r}_{i}$ denotes the results of the races. $I^{r}_{i}=1 (0)$ implies
 that horse $i$ wins (loses) in race $r$. A typical sample from the data is shown
in Table \ref{tab:TS}.

\begin{table}[htbp]
\caption{\label{tab:TS}
Time series of odds and pool for a race that starts at $13:00$.
$N^{r}=10,K^{r}=52$. We show the data only for the first three horses
$1\le i \le 3$.
The first horse wins the race $(I^{r}_{1}=1,I^{r}_{2}=I^{r}_{3}=\cdots=0)$.}
\begin{tabular}{ccccccc}
k & $T^{r}_{k}$ [min] 
& $V^{r}_{k}$ & $O^{r}_{1,k}$ & $O^{r}_{2,k}$& $O^{r}_{3,k}$ & $\cdots$ \\
\hline
1  & 358 &       1  &    0.0 &  0.0 &    0.0 & $\cdots$  \\  
2 & 351  &      169 &   1.6 &  33.3 &  7.9 &  $\cdots$ \\   
3 & 343  &      314 &   1.8 & 11.3 &  8.0 & $\cdots$ \\
4 & 336  &      812  &  2.9 & 17.8 &  14.6 & $\cdots$  \\
5 & 329  &      1400 &  3.3 &  8.6 & 10.6 & $\cdots$ \\
6 & 322  &      1587 &  2.7 &  9.2 &  11.3 & $\cdots$  \\
$\vdots$ & $\vdots$& $\vdots$& $\vdots$& $\vdots$& $\vdots$ & $\vdots$
 \\
51 & 10  &  80064 &   2.4 &  6.4  & 13.4 & $\cdots$ \\
52 & 4 &  148289 &  2.4 &  4.9 & 16.1 & $\cdots$ \\
53 & -2 &  211653 &  2.4 &  5.3  & 17.0 & $\cdots$   
\end{tabular}
\end{table}

 From $O^{r}_{i,k}$, we estimate the win bet fraction $x^{r}_{i,k}$ by
 the following relation according to the rule by JRA.
\begin{equation}
x^{r}_{i,k}=\frac{0.788}{O^{r}_{i,k}-0.1}.
\end{equation}
If the sum of the above values does not equal 1 in each announcement, we renormalize it as 
$\hat{x}^{r}_{i,k}
=\frac{x^{r}_{i,k}}{\sum_{i=1}^{N^{r}}x^{r}_{i,k}}$. Hereafter,
we use $x^{r}_{i,k}$ in place of  $\hat{x}^{r}_{i,k}$.

We use the public win pool averaged over all the races 
as the time variable $t$ for the entire betting process.
For each $0\le v\le 3\times 10^{5}$, we select the nearest $V^{r}_{i,k}$ and
use the average value of $V^{r}_{i,k}$ as the time variable $t$. 
More explicitly, we define $t$ as
\begin{eqnarray}
&&t(v) \equiv \frac{1}{N}\sum_{r=1}^{R} V^{r}_{k^{r}(v)}\cdot N^{r}, \\
&& k^{r}(v)\equiv \{k \hspace*{0.1cm}|\hspace*{0.1cm} 
\mbox{Min}_{k}|V^{r}_{k}-v|  \}.
\end{eqnarray}
The range of $t$ is $70 \le t \le 1.89\times 10^{5}$; the 
largest time is denoted as $t_{f}\equiv 1.89\times 10^{5}$. At $t_{f}$, 
the number of votes $v$ becomes $V^{r}$; $t_{f}$ represents the end of 
the voting process.
If $t$ exceeds $10^{5}$, it cannot be accurately regarded as a time variable.
As the public win pool $V^{r}$ is in the range $10^{5}--3\times 10^{5}$,
if $t$ exceeds the $V^{r}$ of some races, then the voting ends for those races. 
The results of the data analysis for $t\ge 10^{5}$ does not provide 
true information about the time evolution of the voting process. 
The data for $t\ge 10^{5}$ is provided only for the purpose of reference.
We also define $x^{r}_{i}(t)$ as
\begin{equation}
x^{r}_{i}(t)\equiv  x^{r}_{i,k^{r}(v)}.
\end{equation}
The average value of $T^{r}_{k}$ is denoted as  $T(t)$ and defined as 
\begin{equation}
T(t)\equiv \frac{1}{N}\sum_{r=1}^{R}T^{r}_{k^{r}(v)}\cdot N^{r}.
\end{equation}
 
\begin{figure}[htbp]
\begin{center}
\includegraphics[width=8cm,clip]{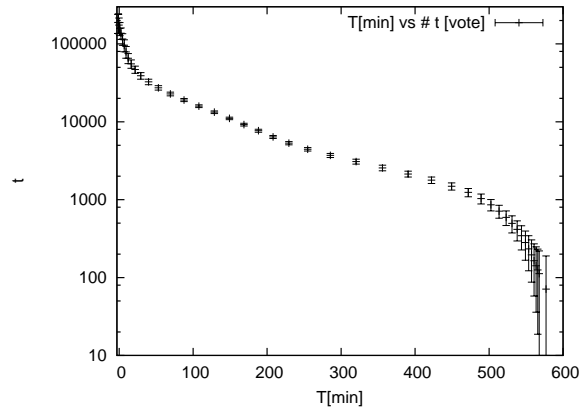}
\end{center}
\caption{Relationship between $T$[min] 
and average number of votes $t$.
In the first announcement ($k=1$), which occurs
approximately 10 h before the start of the race,
the average number of votes $t$ is 71.0. 
Approximately 30 min before the start of the races, about $4\times 10^{4}$ 
votes have been cast.
}
\label{fig:Time_vs_Votes}
\end{figure}

Figure \ref{fig:Time_vs_Votes} shows the relationship between 
$T$ and $t$.
A rapid growth is observed in the average number of
votes $t$ as we approach the start of the race ($T\to 0$). 
Almost half of the votes are thrown in the last 9 min.

In order to provide a  pictorial representation of the betting 
process pictorially, we arrange the 
$N$ horses in the order of the size of $x^{r}_{i}(t)$. We denote the arranged
win bet fraction as $x_{\alpha}(t),\alpha\in \{1,2,\cdots,N\}$.
\begin{equation}
x_{1}(t)\ge x_{2}(t)\ge x_{3}(t)\ge \cdots \ge x_{N}(t) 
\end{equation}
$I_{\alpha}(t)$ tells us  whether horse $\alpha$ wins ($I_{\alpha}=1$) or loses
($I_{\alpha}=0$). In general, the probability that the horse with a large
$x_{\alpha}(t)$ wins is big and vice versa.
We arrange the horses in the 
increasing order of  $\alpha$ from left to right. 
The left-hand side of the sequence represents stronger horses, and  
 the right-hand side, the 
weaker horses. If the win bet fraction does not contain any information
about the strength of the horses, $I_{\alpha}(t)$ randomly assumes the value 
of one and zro. Conversely, if the information is completely
correct, the first $N_{1}$ horses' $I_{\alpha}(t)$ are 1 and the remaining 
$N_{0}$ horses' $I_{\alpha}(t)$ are 0.  
In general, as $t$ increases, the accuracy of $x_{\alpha}(t)$ increases
 and the strong horses with large winning probabilities move to the left 
and vice versa. 

\begin{figure}[htbp]
\begin{center}
\includegraphics[width=12.0cm,clip]{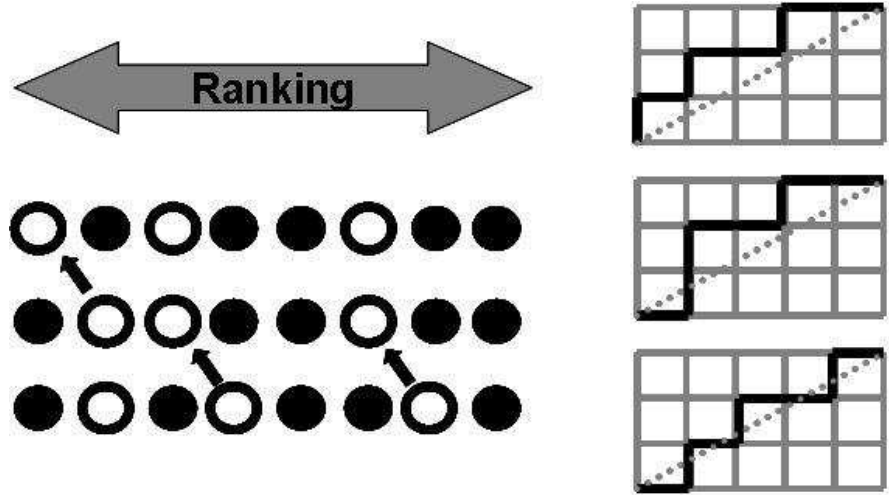}
\end{center}
\caption{Pictorial representation  of the movement of the ranking of the horses by 
betting process. Three horses are winning and 5 horses are losing.
As $t$ increases, we move from the bottom row to the top one.
We depict the winning (losing) horses by white (black) circles.
On the right, we show the corresponding ROC curve.
}
\label{fig:t_vs_Move}
\end{figure}

Figure \ref{fig:t_vs_Move}(left) shows the movement of the ranking of the 
horses due to voting. 
There are eight horses; three of them are winning ones
(white circle) and the remaining five are losing ones (black circle).
The initial configuration, which corresponds to the first announcement $k=1$ 
in each race, is shown in the bottom row. In terms of ranking,
the 2nd, 4th, and 7th horses are winning and the remaining horses are losing
$(I_{2}=I_{4}=I_{7}=1,I_{1}=I_{3}=I_{5}=I_{6}=I_{8}=0)$.
As $t$ increases,
the rank of the winning horses moves to the left in general and 
the accuracy of the prediction by the betters is improved.
At the final state, the 1st, 3rd, and 6th horses are winning ones
$(I_{1}=I_{3}=I_{6}=1,I_{2}=I_{4}=I_{5}=I_{7}=I_{8}=0)$. 

We employ the receiver operating characteristic (ROC) curve \cite{Enleman}
and observe the movement of the ranking  and the 
increase in the accuracy more pictorially. 
It is a path $\{(x_{0,k},x_{1,k})\}_{k=0,\cdots,N}$ in two-dimensional space 
$(x_{0},x_{1})$ from $(x_{0,0},x_{1,0})=(0,0)$ to
$(x_{0,N},x_{1,N})=(1,1)$ as
\begin{equation}
x_{\mu,k}=\frac{1}{N_{\mu}}\sum_{j=1}^{k}\delta_{I_j,\mu}.  
\end{equation}
If $I_{k}=\mu$, the path extends  in $x_{\mu}$ direction.  
If the winning and losing horses are sufficiently mixed in the 
ranking space, the path almost runs diagonally to the end point. 
If the accuracy of the prediction is good, they are separated
 and the path resembles an upward convex curve from 
$(0,0)$ to $(1,1)$.
Figure \ref{fig:t_vs_Move}(right) shows the 
ROC curves corresponding to the ranking configuration 
$\{I_{\alpha}\}_{\alpha=1,\cdots,N}$ on the left. For the bottom case, the win bet 
fractions $x_{\alpha}$
do not contain sufficient information about the strength of the horses.
The ROC curve is almost along the diagonal line.
As the betting progresses from bottom to top, 
the phase separation between the two categories
of the horses does occur and  ROC curves become more and more upwardly convex.

We are able to discuss the discriminative power of the better 
on the basis of the probability 
that the ranking of a randomly selected winning horse $\alpha_{w}$ is higher than that
of a randomly selected losing one $\alpha_{l}$\cite{Enleman}.
The normalized index called 
accuracy ratio AR is defined as
\begin{equation}
\mbox{AR}\equiv 2\cdot (\mbox{Prob}(\alpha_{w}<\alpha_{l})-\frac{1}{2}). 
\end{equation}
If the betters cannot make any discrmination, the horses are
mixed randomly. Prob.$(\alpha_{w}<\alpha_{l})$ becomes  $\frac{1}{2}$ and AR 
becomes zero.
If the betters can make a strong ( or a completely accurate) 
discrimination, both Prob.$(\alpha_{w}>\alpha_{l})$ and AR 
become $1$. 
Prob.$(\alpha_{w}<\alpha_{l})$ is also the area below the 
 ROC curve,  and AR can be estimated as
\begin{equation}
\mbox{AR}=2\cdot \left( \sum_{k=1}^{N}x_{1,k}\cdot (x_{0,k}-x_{0,k-1})
-\frac{1}{2}\right).
\end{equation}
AR changes from $1/15$ to $5/15$ to $7/15$ 
from the bottom row to the top row in Fig \ref{fig:t_vs_Move}.

\section{Power Law Convergence of $x_{\alpha}(t)$ and AR$(t)$} 
\label{Result}

In this section, we explain the results of the analysis of the time series.
We start from the convergence of the win bet fraction $x_{\alpha}(t)$ to its final 
value $x_{\alpha,f}$, where $x_{\alpha,f}$ is the final value of the win bet fraction 
$x_{\alpha,f}\equiv x_{\alpha}(t_{f})$. 
$x_{\alpha,f}$ is the 
winning probability of the horse $\alpha$ 
agreeded by all the betters who vote in the race. 
It is the subjective 
probability or the risk neutral probability of the victory of the horse.
We cannot compute the true winning probability (objective winning probability) of 
the horse. If several horses with almost the same value of the win bet 
fraction $x_{f}$ are grouped together, the winning rate of the horses 
coincides with the win bet 
fraction \cite{Ziemba,Mori2}. 
In this manner, the market is shown to be efficient and no one can get surplus
gain by knowing the value of $x_{f}$. 

We calculate the average value of the squared fluctuation over the horses as
\begin{equation}
[(x_{\alpha}(t)-x_{\alpha,f})^{2}]
\equiv \frac{1}{N}\sum_{\alpha=1}^{N}(x_{\alpha}(t)-x_{\alpha,f})^{2}.
\end{equation}
If the voting has been performed by voters independently, 
$[(x_{\alpha}(t)-x_{\alpha,f})^{2}]$
depends on $t$ as $t^{-1}$; this is termed as normal diffusion. If the behaviour 
deviates from 
$t^{-1}$ to $t^{-\beta}$ with $\beta<1$, the power law convergence is termed as 
 super diffusive \cite{Hod}.  
\begin{figure}[htbp]
\begin{center}
\includegraphics[width=12.0cm,clip]{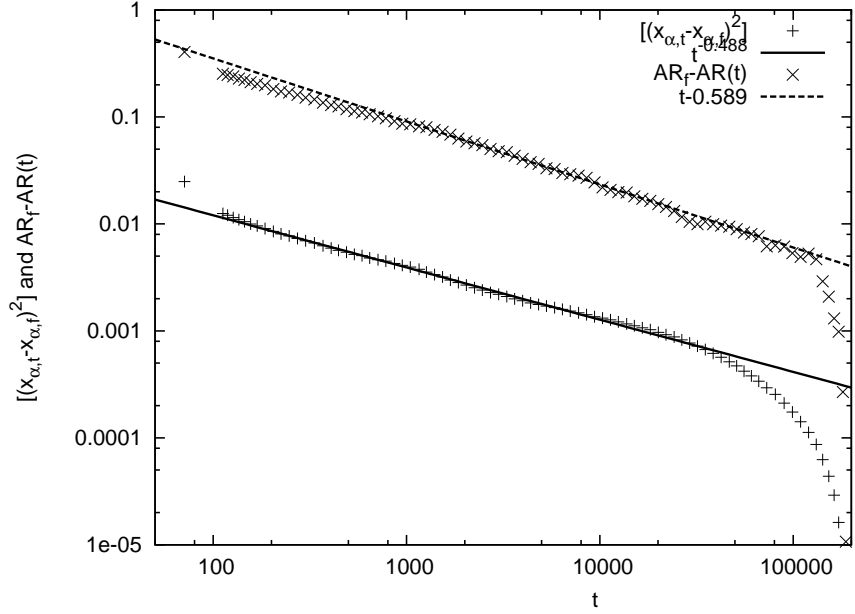}
\end{center}
\caption{Double logarithmic plot of 
$[(x_{\alpha}(t)-x_{\alpha,f})^{2}]$ and AR$_{f}-$AR$(t)$ as functions of  $t$.
The fitted lines with the power law function $a \cdot t^{-\beta}$ are also plotted.}
\label{fig:t_vs_var_dAR}
\end{figure}

Figure \ref{fig:t_vs_var_dAR} shows the double logarithmic plot of 
$[(x_{\alpha}(t)-x_{\alpha,f})^{2}]$ as a function of $t$. 
We observe a slow convergence of  $x_{\alpha}(t)$ to $x_{\alpha,f}$ and 
 a power law behaviour as $t^{-\beta}$ with $\beta=0.488\pm 0.007$ for
$t \le t_{c} \simeq 3\times 10^{4}$.  After $t_{c}$, the convergence occurs rapidly. 
This sort of super diffusive behaviour has been observed in many types of data 
such as coarse-grained DNA sequences, written texts, and financial data. 
Figure \ref{fig:t_vs_var_dAR} also shows the plot of AR$_{f}$-AR$(t)$ as a function of $t$, where AR$_{f}$ is the final value of AR$_{f}\equiv$AR$(t_{f})$.  The fitted curve 
AR$_{f}-a\cdot t^{-\gamma}$ with $\gamma=0.589\pm 0.005$ and 
AR$_{f}=0.6826$ is also shown.
We observe a slow convergence and power law behaviour of AR$(t)$.
Contrary to the convergence of $x_{\alpha}(t)$, the power law relation holds for a  
wider range of $t$. We note that after $t=10^{5}$, the voting ends in some races 
with $V^{r}<t$; in these case
 the plot does not reflect the true time evolution of the voting process.

\section{Voting Model with Independent and Herding Voters} 
\label{Model}

In this section, we introduce a voting model 
and explain the power law behaviours mentioned in the previous section. 
There are  $N^{r}$ horses; this number varies among races.  
We assume it to be a constant value, $N^{r}=N/R$.
Voters vote for the horses individually, and the result of each voting 
is announced promptly. 
The time variable $t\in\{0,1,2\cdots,T\}$ represents 
the number of the votes.
Among $N^{r}$ horses, we select a horse with winning probability
$w$ and call it the target horse. The probability that any other horse 
wins is then $1-w$. Voters somehow
know the value $w$, and after many votes, the win bet fraction
coincides with $w$. We denote the number of votes of the target horse
at time $t$ as $X^{w}_{t}$. At $t=0$, $X^{w}_{t}$ takes the initial value
$X^{w}_{0}=s>0$. There are $N^{r}$ horses in a race and 
the sum of $X_{t}^{w}$ is $N^{r}s+t$. If the target horse gets a vote at $t$,
$X^{w}_{t}$ increases by one unit.
\[
X_{t+1}^{w}=X_{t}^{w}+1. 
\]

\begin{figure}[htbp]
\begin{center}
\includegraphics[width=10.0cm,clip]{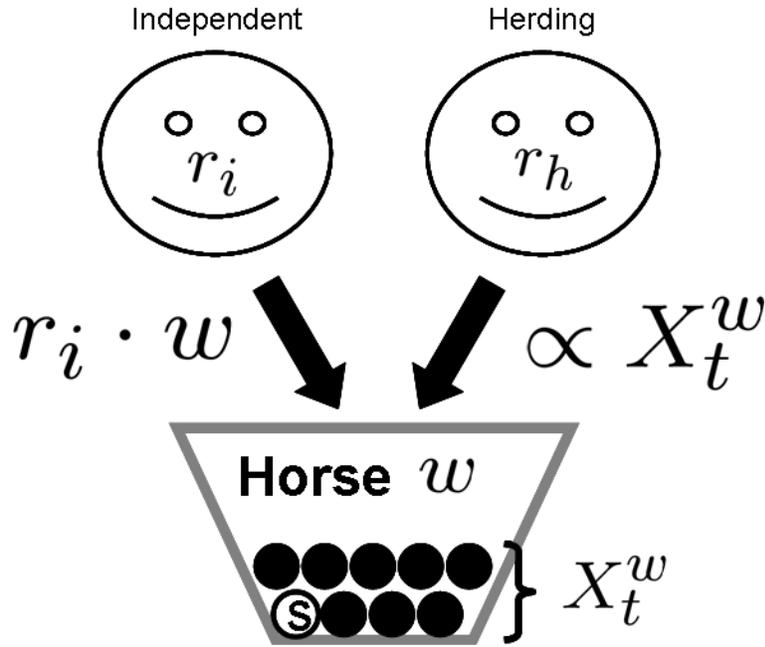}
\end{center}
\caption{Representation of a voting model. There are two types of voters--independent
and  herding. The component rates are $r_{i}$ and $r_{h}$.
The target horse has the winning probability $w$ and the independent
voters voting for it is also $w$. The herding voters decide their vote on 
the basis of the popularity 
$X^{w}_{t}$ of the horse.}
\label{fig:model}
\end{figure}

 We introduce two types of voters--independent and  herding.
 Independent voters their vote on the basis of their private information and are 
 not affected  by the value of $X^{w}_{t}$. These voters provide
 information 
 about the strength of the horses.
 We assume that the ratio of the independent
 voters who vote for the target horse with true winning probability $w$ is $w$. 
  If there are only independent voters in the market, the win bet fraction
 $x_{t}^{w}\equiv \frac{X_{t}^{w}-s}{t}$ converges to the winning probability $w$
 according to the power law $<(x^{w}_{t}-w)^{2}> \sim 1/t$. Here,  
 $<\hspace*{0.3cm}>$ means the averagevalue over all possible paths of the stochastic 
 voting process.
 The herding voters decide their vote on the basis of 
 the popularity of the horse and 
 do not rely on private information. 
 A herding voter casts a vote for the target horse at a rate proportional 
 to $X_{t}^{w}$. They do not bring in any information about the horses. 
 They cause slow convergence of the win bet fraction $x_{t}^{w}$ to the 
 final value $w$ \cite{Hisakado}.
 
 Regarding the rationalities of the voters, we make the following two statements.
 The herding voters are rational because they have no information about the strength 
 of the horses, and the best way for them is to get information from the results 
 of previous votes. If a horse get many votes, many voters agree that the horse 
 is strong. Hence, the herding voters feel it rational to cast a vote to the 
 popular horse and their behaviour is also ratinal.
 Of course, the total votes also include votes cast by herding voters who might  
 provide wrong information. If the ratio of independent voters who vote for  a
 horse with winning probability $w$  
 coincides with $w$, the fluctuation induced by the
 herding voters is cancelled after the votes have been cast 
 many times. 
 
 The independent voters seems  irrational because they vote for the horse 
 they think might win the race.
  In addition, the assumption that the ratio of the independent voters 
 coincides with the 
 true winning probability is unrealistic. 
 However, their voting behaviour  can be understood to be similar to 
 that of fundamental voters. Fundamental voters are rational in the sense that 
 they vote for 
 the horse with the maximum expected payoff. If the win bet fraction is smaller 
 that the true winning 
 probability, the expected payoff of the horse 
 is larger than the average expected payoff of other horses. We assume that the 
 fundamental voters vote for the target horse with the probability 
 proportional to the 
 difference between the win bet fraction and the true winning probability. 
 The resulting 
 probabilistic rule is then transformed to the rule of  the independent 
 and herding voters, as we shall show below. 
 The independent voters can be considered to be rational fundamental voters, 
 if the 
 ratio of the independent voters who vote for the horse with the true winning 
 probability $w$ is equal to $w$.
 
 We denote the component rates of the 
 independent and  herding voters as $r_{i}$ and $r_{h}$, respectively. 
 Obviously, these rates add  up to one, i.e., $r_{i}+r_{h}=1$. 
 Figure \ref{fig:model} explains the model pictorially.
  Mathematically, we can express the above definition of the 
 model using a simple master equation. 
 The probability $P_{t}^{w}$ that the target horse 
 gets a vote  if the horse get $n$ votes up to $t$ and $X^{w}_{t}=n+s$ is
\begin{eqnarray}
P_{t}^{w}(X^{w}_{t}=n+s)&=&r_{i}\cdot w+r_{h}\cdot \frac{n+s}{Z+t},
\\
Z&=&N^{r}\cdot s.
\end{eqnarray}
 The probability $P(n,t+1)$ of finding $n$ votes in $t+1$ voting times 
 follows the evolution equation
\begin{equation}
P(n,t+1)=(1-P_{t}^{w}(n))P(n,t)+P_{t}^{w}(n-1)P(n-1,t).
\end{equation}
If $t$ is sufficiently large and the win bet fraction is $x_{t}^{w}$, 
$P_{t}^{w}$ is expressed as
\begin{equation}
P_{t}^{w}= r_{i}\cdot w+r_{h}\cdot x_{t}^{w}. \label{rule1}
\end{equation}
In the case of fundamental and herding voters, the probabilistic rule is 
expressed for a sufficiently 
large $t$ as
\begin{equation}
P_{t}^{w}=r_{f}\cdot (w+\lambda(w-x_{t}^{w}))+r_{h}'\cdot x_{t}^{w} . \label{rule2}  
\end{equation}
Here, we denote the component ratio of fundamental and herding voters 
as $r_{f}$ and $r_{h}'$, respectively. 
These voters  vote for the target horse with a probability of $w$ if $x_{t}^{w}$ 
 coincides with $w$.
If $x_{t}^{w}$ is different from $w$, the probability of voting for the horse 
changes with $\lambda(w-x_{t}^{w})$
where $\lambda$ is a proportional coefficient.  
By comparing eqs (\ref{rule1}) and (\ref{rule2}), the 
latter model can be mapped to the former model by the relation,
\begin{equation}
r_{i}=(1+\lambda)\cdot r_{f}  \hspace*{0.5cm}\mbox{and}\hspace*{0.5cm}r_{h}=r_{h}'-\lambda r_{f}.
\end{equation}
The voting model of independent and herding voters can be considered to be 
 the voting model of rational voters.

We also assume that $w$ obeys gamma distribution with a shape exponent $a$ and
scale parameter $c$ as
$p_{a,c}(w)$.
\begin{equation}
p_{a,c}(w)\equiv \frac{1}{c\Gamma(a)}(\frac{w}{c})^{s-1}\exp(-w/c).
\end{equation}
The expected value of $w$ is $[w]_{w}=c\cdot a$; we take this value to be $1/N^{r}$.
Here, $[A]_{w}$ is defined as the average of $A(w)$ over 
$p_{a,c}(w)$ as $[A]_{w}\equiv \int_{0}^{1} p_{a,c}(w)A(w)dw$.
In addition, we fit the distribution of $x_{f}$ with the gamma distribution
$p_{a,c}(w)$ by the least square method and we set the value of $a=0.47$ \cite{Mori}. 
The resultant  AR in the model converges to AR$_{f}=0.682$. 
Hereafter, we concentrate on the power law behaviours of the model, 
which do not depend on the detailed nature of the distribution of $w$. 

As has been discussed previously \cite{Hisakado}, the win bet fraction
$x_{t}^{w}$ converges to $w$ after infinite times of voting. 
The convergence shows the  power law 
behaviour as
\begin{eqnarray}
(x_{t}^{w}-w)^{2}\sim t^{-1}\hspace{1cm}\textrm{if}\hspace{0.5cm}r_{i}>\frac{1}{2} \\
(x_{t}^{w}-w)^{2}\sim t^{-2r_{i}}\hspace{1cm}\textrm{if}\hspace{0.5cm}r_{i}<\frac{1}{2} \\
(x_{t}^{w}-w)^{2}\sim \frac{\textrm{log}(t)}{t}\hspace{1cm}\textrm{if}\hspace{0.5cm}r_{i}=\frac{1}{2}.
\end{eqnarray}
The power law exponent does not depend on $w$. After averaging $(x_{t}^{w}-w)^{2}$
over $w$ with $p_{a,c}(w)$, the critical behaviour remains the same. 
We obtain the exponent $\beta$ for the convergence as $\beta=0.488$ and  
we take $r_{i}$
to be half of $\beta$, i.e., as $r_{i}=\beta/2=0.244$. 
The only remaining parameter to be set in the voting model is $s$ the initial 
seeds, i.e., $X_{0}^{w}=s$. This parameter describes the strength of the correlation 
between the votes. In the case  where $r_{i}=0$, the correlation 
function is calculated as $1/(Z+1)=1/(N^{r}s+1)$ \cite{Mori}. If $s$ is small, the votes are concentrated to 
particular horses and the variance of the win bet fraction becomes large.

\begin{figure}[htbp]
\begin{center}
\includegraphics[width=10.0cm,clip]{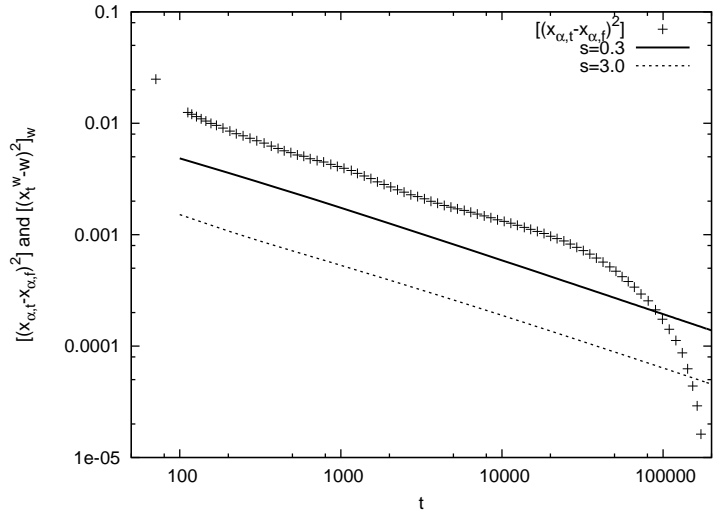}
\end{center}
\caption{
Double logarithmic plot of the averaged win bet fraction
$[(x_{t}^{w}-w)^{2}]_{w}$vs $t$.  
From bottom, we set $s=3$ (dotted) and $0.3$ (solid).
The top line is the plot $[(x_{t}^{w}-w)^{2}]$vs $t$.  
}
\label{fig:var_data_vs_model}
\end{figure}

Figure \ref{fig:var_data_vs_model} shows the result of the convergence 
of $x^{w}(t)$ to $x_{f}$. In the figure, we show the double logarithmic plot 
of $[<(x^{w}(t)-w)^{2}>]_{w}$ vs $t$. We set $s=3$ and $0.3$.
We also show the plot of  
$[(x_{\alpha}(t)-x_{\alpha,f})^{2}]$ for comparison. The two former curves  
are straight lines with a slope of $2r_{i}=\beta$. Compared to the data plot, 
the model's curves show power law behaviour up to the end. We also see a large 
 discrepancy 
between the two plots and the data plot. 
The variance $[(x_{\alpha}(t)-x_{\alpha,f})^{2}]$ is far larger 
than $[<(x^{w}(t)-w)^{2}>]_{w}$. By selecting a small $s$, we can increase 
the variance of 
$x_{t}^{w}$ and  reduce the discrepancy. 
The votes up to the first announcement $k=1$ are highly concentrated
to one or two horses. AR is very small at $k=1$, which can be seen by the presence of
 isolated data points at $t\simeq 70$. The votes provide 
almost no information up to the first announcement ($k=1$). 
The effect of the misleading and concentrated votes
remains during the power law convergence period. It is only after $t_{c}$ that 
the bias begins to disappear quickly.

\begin{figure}[htbp]
\begin{center}
\includegraphics[width=10.0cm,clip]{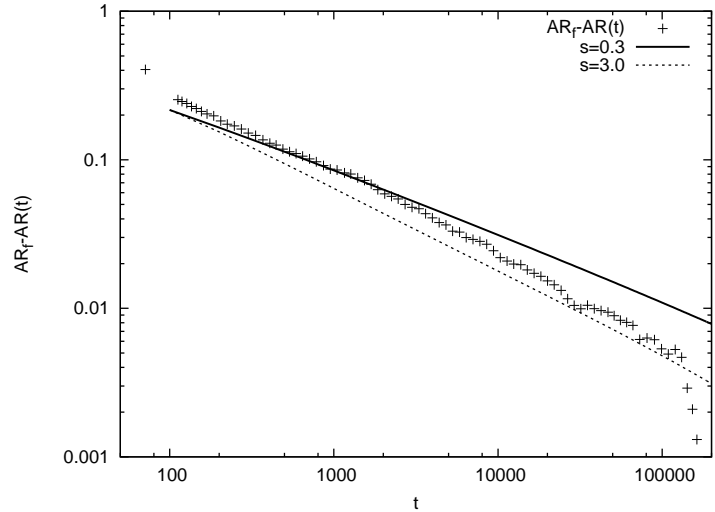}
\end{center}
\caption{
Double logarithmic plot of the averaged win bet fraction
$[(x_{t}^{w}-w)^{2}]_{w}$vs $t$.   
As in the previous figure, we set $s=3$(dotted) and $0.3$(solid).
The top line is the plot $\mbox{AR}_{f}-\mbox{AR}(t)$ vs $t$.  
}
\label{fig:dar_data_vs_model}
\end{figure}

Figure \ref{fig:dar_data_vs_model} shows the result of the 
convergence of AR$(t)$ to AR$_{f}$
 for the same set of the values of $s$.
The double logarithmic plot is straight for $10^{2}\le t \le 2\times 10^{5}$,  and 
the convergence seems to show power law behaviour. Contrary to the convergence of 
$x(t)$, 
 the slope or the critical exponent of the convergence of AR 
depends on $s$. $s=3$ is a
 good selection, as observed from the comparison with the curve of data plot.

\section{Concluding Remarks}
\label{Conclusion}

In this paper, we study the time series data of the win bet in JRA. 
We use the number of votes as the time variable $t$ of the voting process.
 As functions of $t$, the win bet fraction 
$x_{\alpha}(t)$ and the accuracy of predictions 
 AR obey power laws. We obtain $[(x_{\alpha}(t)-x_{\alpha,f})^{2}] 
\sim t^{-0.488}$ and $\mbox{AR}_{f}-
\mbox{AR}(t) \sim t^{-0.589}$. The range where the power law holds is wider in 
AR$(t)$ than in $x_{\alpha}(t)$.
After $t_{c}$, the convergence of the win bet fraction becomes fast. However, 
 AR obey the  power law relation even after $t_{c}$, almost up to $10^{5}$. 

We introduce a simple voting model with two types of voters--independent 
and herding. 
The former voters provide  information on the strengths of the horses 
in the racetrack 
betting market, while latter decide on which horse to 
vote on the basis of the popularities of the horses.
We assume that the component ratio of 
the independent voters coincides with the true winning probability of the
 horse that they vote for and that all the independent voters make rational
decisions.  
We also discuss the relationship between 
the model with independet and herding voters and the 
model with fundamental and herding voters. 
We show that the voting model 
can be used to explain the abovementioned power law behaviours. 
From the exponent of the convergence of the win bet fraction, 
it is observed that
the component ratio of the independent voter to the herding voter is 1:3.

  The change in the convergence after $t_{c}$ cannot be easily explained 
  by increasing the component ratio of independent voters. If this ratio is 
  increased, the win bet 
  fraction converges more rapidly. 
  However, this increase also causes the rapid convergence of AR, which
  contradicts with 
  the  behaviour of the power law convergence of AR even after $t_{c}$. 
  After $t_{c}$, the behaviour of betters can possibly change. 
We believe that the power law of AR holds even after $t_{c}$ because 
the component ratio of the voters who provide 
  information to the market does not change. 
A more detailed analysis of the time series of the 
  betting processes must be performed in the future. 
 In addition, it is also important to 
  study the time dependence of AR. In this study, we have 
numerically analysed AR and showed that it increases
  very slowly and seems to obey the power law relation. 
 In contrast to the win bet fraction, the power law 
  dependence has not been investigated mathematically.
  AR is related to the area under the ROC curve. 
  Statistical properties of the  probabilistic growth of the 
  curve induced by voting  are an interesting 
  problem and we believe that it should be studied.

\section*{Acknowledgment}
This work was supported by a Grant-in-Aid for Challenging 
Exploratory Research 21654054 (SM).

\end{document}